\documentclass[nohyper]{JHEP3}
\usepackage{cite,epsfig,amssymb,amsmath}

\newcommand{\mycomm}[1]{\hfill\break $\phantom{a}$\kern-3.5em{\tt===$>$ \bf #1}\hfill\break}
\newcommand{\mycommA}[1]{\hfill\break $\phantom{a}$\kern-3.5em{\tt   $>$ \bf #1}\hfill\break}

\newcommand{\be}{\begin{equation}}
\newcommand{\ee}{\end{equation}}
\newcommand{\ba}{\begin{eqnarray}}
\newcommand{\ea}{\end{eqnarray}}

\catcode`\@=11 
\def\lsim{\mathrel{\mathpalette\@versim<}}
\def\gsim{\mathrel{\mathpalette\@versim>}}
\def\@versim#1#2{\vcenter{\offinterlineskip
        \ialign{$\m@th#1\hfil##\hfil$\crcr#2\crcr\sim\crcr } }}
\catcode`\@=12 

\title{On threshold resummation of the  longitudinal structure function}

\author{G. Grunberg\\
        Centre de Physique Th\'eorique,  \'Ecole
Polytechnique\\
        91128 Palaiseau Cedex, France\\
        E-mail: \email{georges.grunberg@pascal.cpht.polytechnique.fr}}

\abstract{The validity  of a previously proposed momentum space ansatz for threshold resummation of the non-singlet longitudinal structure function $F_L$ is checked against existing  finite order three-loop results. It is found that the ansatz, which is an assumption for the large-$x$ behavior of the physical evolution kernel, does not work beyond the leading logarithmic contributions to the kernel even at  ${\cal O}(1/(1-x))$ order (except at large-$\beta_0$), which is consistent with a recent observation of Moch and Vogt. 
 Corrections down by one power of $1-x$ are also studied. At ${\cal O}((1-x)^0)$ order, the corresponding ansatz fails already at the leading logarithmic level, where the situation appears similar to that encountered in the case of the $F_i$ ($i=1,2,3$) structure functions. At the next-to-leading logarithmic level,  the same term (with opposite sign) responsible for the failure of the ansatz at ${\cal O}(1/(1-x))$ order is found to  occur.}


\preprint{ }

\begin{document}

\section{Introduction}
Threshold resummation, which deals with the resummation to all orders of perturbation theory of the large logarithmic
corrections  arising from the incomplete cancellation of  soft and collinear gluons at the edge of phase
space, is by now a well developed topic \cite{Sterman:1986aj,Catani:1989ne} in perturbative QCD. Latter, the subject was extended \cite{Akhoury:1998gs,Sotiropoulos:1999hy,Akhoury:2003fw} to cover also the resummation of logarithmically enhanced terms which are  suppressed by some power of the gluon energy $(1-x)$ for $x\rightarrow 1$ in momentum space (or by some power of $1/N$, $N\rightarrow\infty$ in moment space), concentrating on the case of the longitudinal structure function $F_L$ in Deep Inelastic Scattering (DIS) where these corrections are actually the leading terms. Some renewed  interest has been expressed recently in this subject \cite{Kramer:1996iq,Grunberg:2007nc,Laenen:2008ux,Laenen:2008gt,Moch:2009mu,Grunberg:2009yi,Moch:2009hr}. In particular, in \cite{Grunberg:2007nc} a very simple form was obtained  
for the  structure of threshold resummation at all orders in $(1-x)$ in the large--$\beta_0$ limit in {\em momentum space} for the $F_2$ structure function, and a straightforward generalization of the large--$\beta_0$ result to finite $\beta_0$ was suggested, and further developed in \cite{Grunberg:2009yi}. The ansatz in \cite{Grunberg:2007nc,Grunberg:2009yi} was obtained by working at the level of the  momentum space physical evolution kernels (or `physical anomalous dimensions', see e.g. \cite{Furmanski:1981cw,Grunberg:1982fw,Catani:1996sc,Blumlein:2000wh,vanNeerven:2001pe}), which are infrared and collinear safe quantities describing the physical scaling violation, where the structure of the result appears to be particularly transparent. Moreover an ansatz for the  momentum space physical evolution kernel  in the case of  $F_L$, similar to that for $F_2$, was suggested in  \cite{Grunberg:2007nc}. The purpose of this paper is to check the latter ansatz  using the exact three-loop calculations of the longitudinal structure function of \cite{Moch:2004xu,Vermaseren:2005qc}, and to  compare with the closely related approach of \cite{Moch:2009mu}. At leading order in $1-x$,  it is found  that the conjecture  (section 2)  proposed in \cite{Grunberg:2007nc} does not actually  work beyond leading logarithms (section 3), except at large--$\beta_0$, which is consistent with a recent observation of  \cite{Moch:2009mu}. In section 4, an extended ansatz is investigated at the next  order in $1-x$, where it is found to fail already at the leading logarithmic level, as in the case of $F_2$. Concluding remarks are given in section 5, and more technical details are exposed
in four appendix. In particular, the large--$\beta_0$ case is treated extensively in appendix D.

\section{The ansatz}

The scale--dependence of the  deep inelastic ``coefficient functions''
${\cal C}_{a=2,L}(x, Q^2,\mu^2_F)$ corresponding to the flavor  non-singlet $F_a(x,Q^2)$ structure functions ($F_a(x,Q^2)/x={\cal C}_a(x, Q^2,\mu^2_F)\otimes q_{a,ns}(x,\mu^2_F)$, where $ q_{a,ns}(x,\mu^2_F)$ is the corresponding quark distribution)
 can be expressed in terms of ${\cal C}_a(x, Q^2,\mu^2_F)$ itself, yielding the following evolution equation
(see e.g.~Refs.~\cite{Grunberg:1982fw,Catani:1996sc,vanNeerven:2001pe}):
\begin{equation}
\label{F_2_evolution}
\frac{d{\cal C}_a(x,Q^2,\mu^2_F)}{d\ln Q^2}\,=\,\int_x^1 \frac{dz}{z}\,K_a(x/z,Q^2)\,{\cal C}_a(z,Q^2,\mu^2_F)\ ,
\end{equation}
where $\mu_F$ is the factorization scale (we assume for definitness the $\overline{MS}$ factorization scheme is used).
$K_a(x,Q^2)$ is the momentum space \emph{physical evolution kernel}, or {\em physical anomalous dimension}; it is independent of the factorization scale and renormalization--scheme invariant. In \cite{Gardi:2007ma}, using standard results \cite{Sterman:1986aj,Catani:1989ne} of Sudakov resummation in moment space, the result for the leading contribution to this quantity in the $x\rightarrow 1$ limit was derived for $a=2$ (we supress the subscript $a=2$ for simplicity in this case):
\begin{equation}
\label{K_x}
K(x,Q^2)\sim \frac{{\cal J}\left((1-x)Q^2\right)}{1-x}\,+\,\frac{d\ln \left({\cal F}(Q^2)\right)^2}{d\ln Q^2}\,\delta(1-x)\ , \end{equation}
where  ${\cal F}(Q^2)$ is the the quark form factor, and ${\cal J}(Q^2)$, the `physical Sudakov anomalous dimension' (a renormalization sheme invariant quantity), is given by:

\begin{eqnarray} \label{standard-J-coupling}
{\cal J}(Q^2)&= &A\left(a_s(Q^2)\right)+{dB\left(a_s(Q^2)\right)\over d\ln k^2}\\
&=&A\left(a_s(Q^2)\right)+\beta\left(a_s(Q^2)\right)\frac{dB\left(a_s(Q^2)\right)}{da_s}\nonumber\ ,\end{eqnarray}
where
\begin{equation}\label{cusp}
A(a_s)=\sum_{i=1}^\infty
A_ia_s^{i}
\end{equation} 
 is the universal ``cusp'' anomalous dimension \cite{Korchemsky:1993uz} (with $a_s\equiv\frac{\alpha_s}{4\pi}$  the $\overline{MS}$ coupling), 
 
 \begin{equation}\beta(a_s)=\frac{d a_s}{d\ln Q^2}=-\beta_0\, a_s^2-\beta_1\, a_s^3-\beta_2\, a_s^4+...\label{beta}\end{equation}
is the beta function (with $\beta_0=\frac{11}{3}C_A-\frac{2}{3}n_f$)
and
\begin{equation}\label{B-stan}
B(a_s)=
\sum_{i=1}^\infty B_i a_s^{i}
\end{equation}
is the standard final state ``jet function''  anomalous dimension.  It should be noted that both $A(a_s)$ and $B(a_s)$ (in contrast to ${\cal J}(Q^2)$) are
renormalization scheme-dependent quantities.
Eq.(\ref{K_x}) shows that threshold resummation takes a very simple form directly in {\em  momentum-space} when dealing with the physical evolution kernel: ${{\cal J}\left((1-x)Q^2\right)}/({1-x})$ is the leading term in the expansion of the physical momentum space  kernel $K(x,Q^2)$ in the \hbox{$x\to 1$} limit with  $(1-x)Q^2$ fixed, and all threshold logarithms are absorbed into the {\em single} scale  $(1-x)Q^2$. The  term  proportional to $\delta(1-x)$ is comprised of purely virtual corrections associated with the quark form factor. This term is infrared divergent, but  the singularity cancels exactly upon integrating over~$x$ with the divergence of the integral of ${{\cal J} \left((1-x)Q^2\right)}/({1-x})$ near $x\to 1$. Indeed eq.(\ref{K_x}) can be written equivalently \cite{Gardi:2007ma} in term of infrared finite quantities   as:

\begin{equation}
\label{K_x-reg}
K(x,Q^2)\sim\left[\frac{{\cal J}
\left((1-x)Q^2\right)}{1-x}\right]_{+} + \left(\underbrace{
\frac{d\ln \left({\cal F}(Q^2)\right)^2}{d\ln Q^2}\,+\,\int_0^{Q^2}\frac{d\mu^2}{\mu^2}{\cal J}(\mu^2)
}_{\text{infrared \,\,finite}}\right)\,\delta(1-x) \,,
\end{equation}
where 
the integration prescription $[\,]_{+}$ is defined by
\begin{equation}
\label{plus}
\int_0^1 dx\, F(x) \left[\frac{{\cal J}
\left((1-x)Q^2\right)}{1-x}\right]_{+}=\int_0^1dx\,
\Big(F(x)-F(1)\Big)\,\left(\frac{{\cal J}
\left((1-x)Q^2\right)}{1-x}\right)\,.
\end{equation}

In \cite{Grunberg:2007nc}, it was suggested that a result similar to eq.(\ref{K_x-reg}) could also be valid for the (non-singlet) longitudinal structure function $F_L(x, Q^2)$ (with a different physical Sudakov anomalous dimension ${\cal J}_L $ and a different coefficient ${\cal B}(Q^2)$ of the $\delta(1-x)$ term), namely, for $x\rightarrow 1$:

\begin{equation}
\label{K_L}
K_L(x,Q^2)\sim\left[\frac{{\cal J}_L
\left((1-x)Q^2\right)}{1-x}\right]_{+}+\,{\cal B}(Q^2)\,\delta(1-x)\ , \end{equation}
where

\begin{eqnarray} \label{JL-coupling}
{\cal J}_L(Q^2)&= &A\left(a_s(Q^2)\right)+{dB_L\left(a_s(Q^2)\right)\over d\ln Q^2}\\
&=&A\left(a_s(Q^2)\right)+\beta\left(a_s(Q^2)\right)\frac{dB_L\left(a_s(Q^2)\right)}{da_s}\nonumber\ ,\end{eqnarray}
with

\begin{equation}\label{BL}
B_L(a_s)=
\sum_{i=1}^\infty B_{L,i} a_s^{i}
\end{equation}
and

\begin{equation}\label{BL-delta}
{\cal B}(Q^2)=
\sum_{i=1}^\infty b^{(i)} a_s^{i}(Q^2)\ .
\end{equation}

\section{Checking the ansatz}

Let us first derive the ${\cal O}(a_s^3)$ exact result.
Starting from the expansion of the moment space coefficient function

\begin{equation}\label{momentspace-coeff}{\widetilde {\cal C}}_L(Q^2,N,\mu^2_F)=\int_0^1dx\, x^{N-1} {\cal C}_L(x, Q^2,\mu^2_F) \end{equation} 
at $\mu^2_F=Q^2$:

\begin{equation}{\widetilde {\cal C}}_L(N,a_s)\equiv{\widetilde {\cal C}}_L(Q^2,N,\mu^2_F=Q^2)=\sum_{i=1}^\infty {\tilde c}_L^{(i)}(N)\, a_s^i\label{ci-tilde}\end{equation}
where $a_s\equiv a_s(Q^2)$,
the moment space ``physical evolution kernel'' 

 \begin{equation}{\widetilde K}_L(Q^2,N)\equiv\int_0^1dx\, x^{N-1} K_L(x, Q^2) ={d\ln{\widetilde {\cal C}}_L(Q^2,N,\mu^2_F)\over d\ln Q^2}\label{K-tilde1}\end{equation} 
is given by:
 
  \begin{equation}{\widetilde K}_L(Q^2,N)=\gamma(N,a_s)+\beta(a_s)\frac{d\ln {\widetilde {\cal C}}_L(N,a_s)}{da_s}\label{K-tilde2}\ ,\end{equation}
 where $\gamma(N,a_s)=\sum_{i=0}^\infty \gamma_i(N) a_s^{i+1}$ is the non-singlet anomalous dimension in Mellin space. Hence:

\begin{eqnarray}{\widetilde K}_L(Q^2,N)&=&[\gamma_0(N)-\beta_0]a_s+[\gamma_1(N)-\beta_1-\beta_0\  {\tilde \Delta}_2(N)]a_s^2\nonumber\\
& &+[\gamma_2(N)-\beta_2-\beta_1\  {\tilde \Delta}_2(N)-\beta_0\ (2 {\tilde \Delta}_3(N)- {\tilde \Delta}_2(N)^2)]a_s^3+...\ ,\label{KL-N-expand}\end{eqnarray}
where

\begin{equation}{\tilde \Delta}_i(N)=\frac{{\tilde c}_L^{(i)}(N)}{{\tilde c}_L^{(1)}(N)}\label{di-N}\ .
\end{equation}
Thus for the momentum space kernel one gets:

\begin{eqnarray}\label{KL-x-expand} K_L(x,Q^2)&=&P_0(x)\ a_s+[P_1(x)-\beta_0\   \Delta_2(x)]a_s^2\nonumber\\
& &+[P_2(x)-\beta_1\  \Delta_2(x)-\beta_0(2 \Delta_3(x)-\Delta_2^{\otimes 2}(x))]a_s^3+...\\
& &-(\beta_0\ a_s+\beta_1\ a_s^2+\beta_2\ a_s^3+...)\delta(1-x)\nonumber
\ ,\end{eqnarray}
where $P_i(x)$ are the standard $(i+1)$-loop splitting functions,

\begin{equation}\Delta_i(x)=-\frac{x^2}{4\ C_F}\frac{d}{d x}\Big(\frac{c_L^{(i)}(x)}{x}\Big)+\delta_i\ \delta(1-x)\label{di-x}\end{equation}
 is the inverse Mellin of  ${\tilde \Delta}_i(N)$, with the delta function coefficients $\delta_i$ to be determined\footnote{$\delta_2$ is given in eq.(\ref{delta2}).} accordingly:

\begin{equation}{\tilde \Delta}_i(N)=\int_0^1dx\, x^{N-1}\Delta_i(x)\ ,\label{di-N-tilde}\end{equation}
and $\Delta_2^{\otimes 2}(x)\equiv\Delta_2(x)\otimes\Delta_2(x)$.
Moreover  $c_L^{(i)}(x)$ is   the inverse Mellin of ${\tilde c}_L^{(i)}(N)$ defined in eq.(\ref{ci-tilde}):

\begin{equation}{\tilde c}_L^{(i)}(N)=\int_0^1dx\, x^{N-1} c_L^{(i)}(x)\ ,\label{ci-moments} \end{equation} 
such that

\begin{equation} {\cal C}_L(x,Q^2,\mu^2_F=Q^2)=\sum_{i=1}^\infty c_L^{(i)}(x)\, a_s^i\ ,\label{ci}\end{equation}
and the input ${\tilde c}_L^{(1)}(N)=4 C_F/(N+1)$ has been used. 

Considering now the $x\rightarrow 1$ limit, the leading ${\cal O}(1/(1-x))$ term in the physical kernel $K_L(x,Q^2)$ to ${\cal O}(a_s^3)$  can be derived from the well-known result \cite{Korchemsky:1988si} for the splitting functions

\begin{equation}P_i(x)\sim \frac{A_{i+1}}{(1-x)_+}+...\label{cusp-1}\ ,\end{equation}
where the $A_i$ are known \cite{Kodaira:1981nh,Moch:2004pa} up to $i=3$, together with the the two \cite{SanchezGuillen:1990iq,vanNeerven:1991nn,Moch:1999eb}  and three loop \cite{Moch:2004xu,Moch:2009hr,Vermaseren:2005qc,Moch:2009mu} results for the longitudinal coefficient function. The result of the exact calculation is most easily stated by comparing it with the result following from the ansatz eq.(\ref{K_L}).

The renormalization group invariance of ${\cal J}_L$ yields the standard relation:

\begin{eqnarray}{\cal J}_L\left((1-x)Q^2\right)&=&j_{L,1}\ a_s+a_s^2[-j_{L,1}\beta_0 L_x+j_{L,2}]\label{jL-expand}\\
&+ &a_s^3[j_{L,1}\beta_0^2 L_x^2-(j_{L,1}\beta_1+2 j_{L,2} \beta_0) L_x+j_{L,3}]\nonumber\\
&+ &a_s^4[-j_{L,1}\beta_0^3 L_x^3+(\frac{5}{2}j_{L,1}\beta_1\beta_0+3 j_{L,2} \beta_0^2) L_x^2-(j_{L,1}\beta_2+2 j_{L,2} \beta_1+3 j_{L,3} \beta_0) L_x+j_{L,4}]+...\ ,\nonumber
\end{eqnarray}
where $a_s=a_s(Q^2)$ and $L_x\equiv \ln(1-x)$.
Comparing eq.(\ref{K_L}) and (\ref{jL-expand}) with the exact result for the leading ${\cal O}(1/(1-x))$ term in $K_L(x,Q^2)$, one finds (see appendices (A-C)):

i) The leading logarithms in the exact result agree with the leading logarithms in  eq.(\ref{jL-expand}), provided

\begin{equation}j_{L,1}=A_1=4C_F\label{jL1}\ .\end{equation}
Eq.(\ref{jL1})   corresponds to eq.(7) in \cite{Moch:2009mu}. I note that this result implies that the $\frac{\ln^3(1-x)}{1-x}$ terms present separately in $\Delta_3(x)$ and $\Delta_2^{\otimes 2}(x)$ cancel in the combination $2 \Delta_3(x)-\Delta_2^{\otimes 2}(x)$.

ii) There is a discrepancy at the next-to-leading logarithmic  level, starting at  ${\cal O}(a_s^3)$, namely:

\noindent Comparing the ${\cal O}(a_s^2)$ term in eq.(\ref{jL-expand}) with the corresponding term in the exact result, one determines

\begin{equation}j_{L,2}=A_2+\beta_0(\beta_0-C_F)+4\beta_0(1-\zeta_2)(C_A-2C_F)\label{jL2}\  ,\end{equation}
with the two-loop cusp anomalous dimension $A_2=(16/3-8\zeta_2)C_F C_A+(20/3)C_F \beta_0$.
Eq.(\ref{jL2}) fixes $B_{L,1}$ (see eq.(\ref{JL-coupling})):
\begin{equation}B_{L,1}=-[\beta_0-C_F+4(1-\zeta_2)(C_A-2C_F)]\label{BL1}\  .\end{equation}
However, comparing with the exact coefficient of the  single $\frac{\ln(1-x)}{1-x}$ term occurring at ${\cal O}(a_s^3)$ in $K_L(x,Q^2)$, one finds the latter is {\em not} given by $-(j_{L,1}\beta_1+2 j_{L,2} \beta_0)\frac{\ln(1-x)}{1-x}$ as predicted by eq.(\ref{K_L}) and (\ref{jL-expand}), but rather by:

\begin{equation}-[j_{L,1}\beta_1+2 (j_{L,2}+\Delta_{L,2}) \beta_0]\frac{\ln(1-x)}{1-x}\label{discrepancy}\ ,\end{equation}
with

\begin{equation}\Delta_{L,2}=-16(C_A-2C_F)^2 (1-3\zeta_2+\zeta_2^2+\zeta_3)\label{DL2}\ .\end{equation}
Eq.(\ref{discrepancy}) and (\ref{DL2}) represent the equivalent of eq.(22) of \cite{Moch:2009mu} at the level of the physical kernel. Furthermore,  $j_{L,3}$  could be extracted from the ${\cal O}(a_s^3)$ non-logarithmic terms, and thus also  $B_{L,2}$, since (eq.(\ref{JL-coupling})):
\begin{equation}j_{L,3}=A_3-\beta_1\ B_{L,1}-2\beta_0\ B_{L,2}\label{B_L2}\ .\end{equation}

\underline{Large $\beta_0$ case}: it shown in appendix D that
 eq.(\ref{K_L}) is indeed valid at large-$\beta_0$, where it takes the more special form:

\begin{equation}
\label{K_L-large-b0}
\left.K_L(x, Q^2)\right\vert_{\rm large \,\,\beta_0}\,\sim\left[\frac{d{\cal B}_{\infty}\left((1-x)Q^2\right)/d\ln Q^2}{1-x}\right]_+\,+\,{\cal B}_{\infty}(Q^2)\,\delta(1-x)\ , \end{equation}
i.e. $\left.{\cal J}_L\left(Q^2\right)\right\vert_{\rm large \,\,\beta_0}\,=\frac{d{\cal B}_{\infty}\left(Q^2\right)}{d\ln Q^2}$ is the  derivative of the large-$\beta_0$ delta function coefficient $\left.{\cal B}\left(Q^2\right)\right\vert_{\rm large \,\,\beta_0}\,={\cal B}_{\infty}(Q^2)$. Collecting all the  $\delta$ function contributions in  eq.(\ref{KL-x-expand}) one easily finds, since the splitting functions contributions can be neglected at large $\beta_0$:

\begin{eqnarray}
\label{B-large-b0}\left.{\cal B}\left(Q^2\right)\right\vert_{\rm large \,\,\beta_0}&=&-\beta_0 a_s-\beta_0  \left.\delta_2\right\vert_{\rm large \,\,\beta_0}a_s^2+{\cal O}(a_s^3)\nonumber\\
&=&-\beta_0 a_s-\frac{19}{6}\beta_0^2 a_s^2+{\cal O}(a_s^3)\ ,
\end{eqnarray}
where I used eq.(\ref{delta2}) to get $\left.\delta_2\right\vert_{\rm large \,\,\beta_0}=\frac{19}{6}\beta_0$, which agrees with the value  $\left. b^{(2)}\right\vert_{\rm large \,\,\beta_0}=-b_{\infty}^{(2)}\beta_0^2=-\frac{19}{6}\beta_0^2$ derived in a different manner in eqs.(\ref{large-b0-coeff}) and (\ref{b-large-b0}).
Comparing eq.(\ref{K_L-large-b0}) with eq.(\ref{JL-coupling}) also shows that $\left.{\cal B}\left(Q^2\right)\right\vert_{\rm large \,\,\beta_0}\,= \left. B_L\left(a_s(Q^2)\right)\right\vert_{\rm large \,\,\beta_0}$.

\section{Threshold resummation beyond leading order}
One can also try to resum large $x\rightarrow 1$ logarithms suppressed by one power of $1-x$ wih respect to the leading terms. Here the situation looks similar to the one prevailing \cite{Grunberg:2009yi,Moch:2009hr} for the $F_i$ ($i=1,2,3$)  structure functions. The exact calculation gives for $x\rightarrow 1$:

\begin{eqnarray}K_L(x,Q^2)&\sim&\frac{1}{1-x}\Big[j_{L,1}\ a_s+a_s^2[-j_{L,1}\beta_0 L_x+j_{L,2}]\label{K_L-NLL2}\\
&+ &a_s^3[j_{L,1}\beta_0^2 L_x^2-(j_{L,1}\beta_1+2( j_{L,2} +\Delta_{L,2})\beta_0) L_x+j_{L,3}]+{\cal O}(a_s^4)\Big]\nonumber\\
&+ &\Big[-j_{L,1}\ a_s+a_s^2(16 C_F^2 L_x+{\cal O}(L_x^0))\nonumber\\
& &+a_s^3(-24\beta_0 C_F^2L_x^2+{\cal O}(L_x))+{\cal O}(a_s^4)\Big]\ ,\nonumber
\end{eqnarray}
where I used that \cite{Floratos:1977au,Curci:1980uw,Kodaira:1981nh,Moch:2004pa}:

\begin{eqnarray}
\label{P1}
P_0(x)&\sim& \frac{A_1}{1-x}-A_1+{\cal O}(1-x)\nonumber\\
P_1(x)&\sim& \frac{A_2}{1-x}+A_1^2ÊL_x+{\cal O}(L_x^0)\nonumber\\
P_2(x)&\sim& \frac{A_3}{1-x}+2A_1A_2ÊL_x+{\cal O}(L_x^0)\ ,\end{eqnarray}
and  the $\delta(1-x)$ terms have been dropped.

\noindent\underline{Leading logarithms}:
considering first the leading logarithms in the last two lines of eq.(\ref{K_L-NLL2}) suggests to try the ansatz\footnote{An expansion in $1/(1-x)$, rather then in $1/r=x/(1-x)$ as in the case of the $F_2$ structure function, appears more appropriate here, leading to a simpler result eq.(\ref{bar-j02L}).} involving a single `explicit' logarithm (ignoring for the moment the `anomalous' next-to-leading logarithmic $\Delta_{L,2}$ contribution to the ${\cal O}(1/(1-x))$ term):

\begin{equation}
\label{K_L-ansatz3}
K_L(x,Q^2)\sim\frac{1}{1-x} {\cal J}_L\left((1-x)Q^2\right)+\bar{{\cal J}}_{0L}\left((1-x)Q^2\right)\,\ln(1-x)-j_{L,1}a_s+{\cal O}(a_s^2)\  ,\end{equation}
where the  ${\cal O}(a_s^2)$ part refers to terms which contribute only to sub-leading logarithms. 
Setting:
\begin{equation}\bar{{\cal J}}_{0L}\left((1-x)Q^2\right)=\bar{j}_{0L,2}a_s^2+a_s^3(-2\bar{j}_{0L,2}\beta_0 L_x)+\bar{j}_{0L,3})+...\label{jbar0L-2loop}\ ,\end{equation}
one must require:

\begin{equation}\bar{j}_{0L,2}=16 C_F^2\label{bar-j02L}\end{equation}
to match   the coefficient of the  $a_s^2 L_x$ term  in $K_L(x,Q^2)$.
The ansatz eq.(\ref{K_L-ansatz3}) then predicts the coefficient of the  $a_s^3 L_x^2$ term to be $- 32 C_F^2\beta_0$, instead of the exact value $- 24 C_F^2\beta_0$. I note however that the color factors are correctly reproduced (as well as the absence of Riemann $\zeta$ function contributions).
The exact  result also shows that  the $L_x^2$ term in the  combination $2 \Delta_3(x)-\Delta_2^{\otimes 2}(x)$ occurring  in eq.(\ref {KL-x-expand}) is contributed only by the $C_F^2$ color factor, which implies the corresponding contributions of the $C_F\beta_0$ and $C_F C_A$ color factors cancel between $\Delta_3(x)$ and $\Delta_2^{\otimes 2}(x)$: this observation allows to relate   $3$-loop coefficients to $2$-loop ones for these color factors.

\noindent Moreover,  both the exact  and the predicted  values of the coefficient of the ${\cal O}(a_s^3)$ $C_F^2\beta_0 L_x^2$ term turn out to be {\em the same}  \cite{Grunberg:2009yi,Moch:2009hr}   for  the physical evolution kernels $K_i$ associated to $F_i$ ($i=1,2,3$)  and for $K_L$.  This equality may well be accidental, since this ${\cal O}(a_s^3)$ coefficient is contributed by $3$-loop coefficient functions in the case of $K_L$, but by $2$-loop ones in the case of $K_i$. If however there is a deeper reason for it, one would predict the ${\cal O}(a_s^4)$  $C_F^2\beta_0^2 L_x^3$ term in $K_L$ (contributed by $4$-loop coefficient functions)  to be given by $\frac{88}{3}C_F^2\beta_0^2L_x^3$, namely by  the corresponding ${\cal O}(a_s^4)$  $C_F^2\beta_0^2 L_x^3$  term \cite{Grunberg:2009yi,Moch:2009hr}  in $K_i$ (contributed by $3$-loop coefficient functions).

\noindent \underline{Next-to-leading logarithms}:
using the  results in \cite{Moch:2009mu}, it is also possible to compute the coefficient of the $a_s^3\,  {\cal O}(L_x)$  term in the last line of eq.(\ref{K_L-NLL2}), which requires in addition the knowledge  of  $d_{31}+c_{31}$ (in the notation of appendix A); the latter information can be derived \cite{Vogt}   from the exact results in \cite{Vermaseren:2005qc}. An interesting fact emerges concerning the `anomalous'
$\Delta_{L,2}$ contribution. In leading $1/(1-x)$ order, it was defined as a discrepancy with respect to the prediction of the  ansatz eq.(\ref{K_L}). It turns out it also coincides with the contribution of the ${\cal O}(a_s^3)$ $\beta_0(C_A-2C_F)^2$ `non-planar' color factor,  if one uses a basis of color factors made of products of the three basic factors $C_F$, $C_A-2C_F$ and $\beta_0$. In this basis, one finds the contribution of the $a_s^3 \beta_0(C_A-2C_F)^2  L_x$ term to be {\em minus} that of the corresponding $a_s^3 \beta_0(C_A-2C_F)^2  L_x/(1-x)$ term. Namely the ${\cal O}(a_s^3)$ $\beta_0(C_A-2C_F)^2$ single logarithmic contribution to $K_L$ is given by (see eq.(\ref{2d3-d2d2})):

\begin{equation} -2\beta_0\Delta_{L,2}\Big(\frac{1}{1-x}-1\Big)\ln(1-x)\label{anomalous}\ .\end{equation}

\section{Conclusion}
The suggestion that the longitudinal physical evolution kernel depends at large $x$ upon the single scale $(1-x)Q^2$ (similarly to the physical kernels $K_i$ associated to the $F_i$ structure functions, $i=1,2,3$), which  is indeed correct in the large-$\beta_0$ limit, is seen to fail at finite $\beta_0$ even in the leading  ${\cal O}(1/(1-x))$ order. While the ansatz is probably sustained to all orders in $a_s$ at the level of the leading logarithmic contributions to the kernel, an obstacle to threshold resummation  appears at  ${\cal O}(a_s^3)$  in the next-to-leading logarithmic contribution. Interestingly, the very same next-to-leading logarithmic term responsible for the problem also shows up in the kernel (but with opposite sign) at  ${\cal O}((1-x)^0)$ order. At the same  ${\cal O}((1-x)^0)$  order,  another obstruction to threshold resummation already appears in $K_L$ at the leading logarithmic level,  which looks more similar to the one  encountered \cite{Grunberg:2009yi,Moch:2009hr}  in the case of the $K_i$'s ( $i=1,2,3$). The results in \cite{Moch:2009hr} make it clear that the success of the present ${\cal O}(1/(1-x))$ ansatz for $K_L$   at the leading logarithmic  level is related to the universality \cite{Grunberg:2009yi,Moch:2009hr} of the leading logarithmic contribution to the  $K_i$'s at ${\cal O}((1-x)^0)$ order. It would be interesting to find out whether the additional relations in $K_L$ observed in the present paper    at  ${\cal O}((1-x)^0)$ order, if not accidental,  could also be linked  to a simple property of the difference $K_1-K_2$.

\vspace{0.5cm}

\noindent\textbf{Acknowledgements}

\vspace{0.3cm}

\noindent I thank Andreas Vogt for communicating the exact results for the $c_{ik}$'s and the $d_{ik}$'s defined in appendix A.

\appendix

\section{Soft part  of $c_L^{(i)}(x)$}
Setting in the limit  $x\rightarrow 1$:

 \begin{equation}c_L^{(i)}(x)=\sum_{k=0}^{2i-2} \ln^k(1-x)[ c_{ik}+(1-x)d_{ik}]\ ,\label{cLi-expand}\end{equation}
we have for $i=1$:

\begin{equation}\label{c1} c_{10}=4 C_F\end{equation}
and:

\begin{equation}\label{d1} d_{10}+c_{10}=0\ .\end{equation}
For $i=2$ one gets:

\begin{eqnarray}\label{c2k}
c_{22}&=& 8 C_F^2\\
c_{21}&=& 4 C_F[C_F -\beta_0-4(1-\zeta_2)(C_A-2 C_F)]\nonumber\\
c_{20}&=& 4 C_F[\frac{19}{6}\beta_0-(\frac{47}{6}+4\zeta_2)C_F+(\frac{1}{3}+4\zeta_2-6\zeta_3)(C_A-2 C_F)]\nonumber\ ,\end{eqnarray}
and:

\begin{eqnarray}\label{d2k}
d_{22}+c_{22}&=& 0\\
d_{21}+c_{21}&=& 32 C_F^2\nonumber\\
d_{20}+c_{20}&=& 4 C_F[3(C_F -\beta_0)-4(1-\zeta_2)(C_A-2 C_F)]\nonumber\ .\end{eqnarray}
For $i=3$, we have \cite{Moch:2009hr,Vermaseren:2005qc}:

\begin{eqnarray}\label{c3k}
c_{34}&=& 8 C_F^3\\
c_{33}&=& 8 C_F^2[C_F -\frac{4}{3}\beta_0-4(1-\zeta_2)(C_A-2 C_F)]\nonumber\\
c_{32}&=&C_F\Big[  (-\frac{118}{3}-96\zeta_2)C_F^2+4\beta_0^2+\frac{98}{3}C_F\beta_0+24(1-\zeta_2)(C_A-2C_F)\beta_0\nonumber\\
& &+(-\frac{8}{3}+32\zeta_2-48\zeta_3)C_F(C_A-2C_F)-32(\zeta_3-\zeta_2)(C_A-2C_F)^2\Big]\nonumber\ ,\end{eqnarray}
 while the following results  can be easily deduced from \cite{Moch:2009mu}:

\begin{eqnarray}\label{d3k}
d_{34}+c_{34}&=&0\\
d_{33}+c_{33}&=& 64 C_F^3\nonumber\\
d_{32}+c_{32}&=& 8 C_F^2[ C_F -9 \beta_0-20(1-\zeta_2)(C_A-2 C_F)]\nonumber\  .\end{eqnarray}
Moreover one can also derive \cite{Vogt} $c_{31}$ and  $d_{31}$ from the results in  \cite{Vermaseren:2005qc}. Here I quote  the  outcome  for  $d_{31}+c_{31}$:

\begin{eqnarray}\label{d31}
d_{31}+c_{31}&=&C_F\Big[  (0\times\zeta_3+0\times\zeta_2^2+...)C_F^2+24\beta_0^2+\frac{500}{3}C_F\beta_0+(240-192\zeta_2)(C_A-2C_F)\beta_0\nonumber\\
& &+(-192\zeta_3+0\times\zeta_2^2+...)C_F(C_A-2C_F)-64(\zeta_3-\zeta_2)(C_A-2C_F)^2\Big]\ ,\end{eqnarray}
where  only the $\zeta_3$-dependent terms  (there is no $\zeta_2^2$ term) have been written down  for  the $C_F^2(C_A-2C_F)$ color factor (the  $C_F^3$ factor has no $\zeta_3$ contribution either).

\section{Soft parts of $\Delta_2(x)$ and $\Delta_2^{\otimes 2}(x)$}

For $x\rightarrow 1$, one obtains using eq.(\ref{di-x}):

\begin{eqnarray}\label{d2-soft}
\Delta_2(x)&=&\frac{1}{4 C_F}\Big[2 c_{22}\frac{\ln(1-x)}{1-x}+c_{21}\frac{1}{1-x}\Big]+\delta_{2}\ \delta(1-x)\\
& &+\frac{1}{4 C_F}\Big[(d_{22}+c_{22})\ln^2(1-x)+(d_{21}+c_{21}+2 d_{22}-2 c_{22})\ln(1-x)+d_{20}+c_{20}+ d_{21}- c_{21}\Big]\nonumber\  ,\end{eqnarray}
where the $\ln^p(1-x)/(1-x)$ terms ($p\geq 0$) should be interpreted as usual as $+$-distributions, which implies that

\begin{equation}\label{delta2}\delta_2=\frac{1}{4 C_F}c_{20}\ ,\end{equation}
with $c_{20}$ given in eq.(\ref{c2k}).
 One thus gets:

\begin{eqnarray}\label{d2-soft-bis}
\Delta_2(x)&=&\frac{1}{1-x}[4 C_F\ln(1-x)+B_{L,1}]+\delta_2\ \delta(1-x)\\
& &+0\times\ln(1-x)+10 C_F-2\beta_0-B_{L,1}\nonumber\  ,\end{eqnarray}
where $B_{L,1}$ is given in eq.(\ref{BL1}). I note that not only the ${\cal O}(\ln^2(1-x))$ term, but also the 
${\cal O}(\ln(1-x))$ term, is absent in eq.(\ref{d2-soft-bis}).

Using e.g. the formulas in Appendix A of \cite{Grunberg:2009yi}, one deduces the soft part of $\Delta_2^{\otimes 2}(x)$:

\begin{eqnarray}\label{d2d2-soft}
\Delta_2^{\otimes 2}(x)&=&a_3\frac{\ln^3(1-x)}{1-x}+a_2\frac{\ln^2(1-x)}{1-x}+a_1\frac{\ln(1-x)}{1-x}+a_0\frac{1}{1-x}\\
& &+b_3\ln^3(1-x)+b_2\ln^2(1-x)+b_1\ln(1-x)+b_0\nonumber\  ,\end{eqnarray}
(where the $\delta(1-x)$ term has been skept), with:

\begin{eqnarray}\label{ai}a_3&=&16 C_F^2\\
a_2&=&12 C_F [-\beta_0+C_F-4(1-\zeta_2)(C_A-2C_F)]\nonumber\\ 
a_1&=&-32\zeta_2 C_F^2+8 C_F\delta_2+2 B_{L,1}^2\nonumber\\
a_0&=&-8\zeta_2 C_F B_{L,1}+32\zeta_3 C_F^2+2 B_{L,1}\delta_2\nonumber\  ,\end{eqnarray}
and:

\begin{eqnarray}\label{bi}b_3&=&0\\
b_2&=&4 C_F[-\beta_0+13 C_F+4(1-\zeta_2)(C_A-2 C_F)]\nonumber\\ 
b_1&=& 2 B_{L,1}(14 C_F-2\beta_0)- 2 B_{L,1}^2\nonumber\  .\end{eqnarray}
Eqs.(\ref{ai}) and (\ref{bi}) yield:

\begin{eqnarray}\label{a1}
a_1&=& (-\frac{182}{3}-64\zeta_2)C_F^2+2\beta_0^2+\frac{64}{3}C_F\beta_0+16(1-\zeta_2)(C_A-2C_F)\beta_0\nonumber\\
& &+(-\frac{40}{3}+48\zeta_2-48\zeta_3)C_F(C_A-2C_F)+32(1-\zeta_2)^2(C_A-2C_F)^2\ ,\end{eqnarray}
and

\begin{eqnarray}\label{b1}
b_1&=& -30C_F^2-6\beta_0^2+36C_F\beta_0-32(1-\zeta_2)(C_A-2C_F)\beta_0\nonumber\\
& &+128(1-\zeta_2)C_F(C_A-2C_F)-32(1-\zeta_2)^2(C_A-2C_F)^2\ .\end{eqnarray}
I note that $b_1$ (contrary to $a_1$) {\em does not have any} $\zeta_3$ contribution: this fact (as well as the vanishing of $b_3$) is due to the absence of a  ${\cal O}(\ln(1-x))$ term in eq.(\ref{d2-soft-bis}).

\section{Soft part of $\Delta_3(x)$ }

For $x\rightarrow 1$,  eq.(\ref{di-x}) yields:

\begin{eqnarray}\label{d3-soft}
\Delta_3(x)&=&\frac{1}{4 C_F}\Big[4 c_{34}\frac{\ln^3(1-x)}{1-x}+3 c_{33}\frac{\ln^2(1-x)}{1-x}+2 c_{32}\frac{\ln(1-x)}{1-x}+c_{31}\frac{1}{1-x}\\
& &+(d_{34}+c_{34})\ln^4(1-x)+(d_{33}+c_{33}+4 d_{34}-4 c_{34})\ln^3(1-x)\nonumber\\
& &+(d_{32}+c_{32}+3 d_{33}-3 c_{33})\ln^2(1-x)+(d_{31}+c_{31}+2 d_{32}-2 c_{32})\ln(1-x)+...\Big]\nonumber\  ,\end{eqnarray}
where the $\delta(1-x)$ term has been skept. One thus gets:

\begin{eqnarray}\label{d3-soft-bis}
\Delta_3(x)&=&\frac{1}{1-x}\Big[8 C_F^2\ln^3(1-x)+C_F\Big(-8 \beta_0+6 C_F-24(1-\zeta_2) (C_A-2C_F)\Big)\ln^2(1-x)\nonumber\\
& &+\Big((-24\zeta_3+0\times\zeta_2^2+...)C_F(C_A-2C_F)-16(\zeta_3-\zeta_2)(C_A-2C_F)^2+...\Big)\ln(1-x)+...\Big]\nonumber\\
& &+\Big[0\times\ln^3(1-x)+2C_F\Big(- \beta_0+19 C_F+4(1-\zeta_2)(C_A-2C_F)\Big)\ln^2(1-x)\\
& &+\Big((0\times\zeta_3+0\times\zeta_2^2+...)C_F(C_A-2C_F)+16(\zeta_3-\zeta_2)(C_A-2 C_F)^2+...\Big)\ln(1-x)+...\Big]\nonumber\  ,\end{eqnarray}
where   only the contribution of the  $(C_A-2 C_F)^2$ color factor, as well as the $\zeta_3$ contribution to the $C_F(C_A-2C_F)$ factor (which has no $\zeta_2^2$ contribution),  have been written down in the coefficients of the $\ln(1-x)/(1-x)$ and\footnote{The $C_F(C_A-2C_F)$ color factor in the  coefficient of the $\ln(1-x)$  term has no $\zeta_3$ contribution either.} $\ln(1-x)$  terms (the other color factors have no $\zeta_3$, nor $\zeta_2^2$,  contributions).   I note that not only the ${\cal O}(\ln^4(1-x))$, but also the ${\cal O}(\ln^3(1-x))$ terms are absent in the ${\cal O}((1-x)^0)$ part of eq.(\ref{d3-soft-bis}). Moreover the $( C_A-2 C_F)^2$ color factor contributions to  the  $\ln(1-x)/(1-x)$ and  $\ln(1-x)$ terms are {\em equal up to a sign}, and the same statement holds for the corresponding contributions in $\Delta_2^{\otimes 2}(x)$, see eq.(\ref{a1}) and (\ref{b1}).

Consider now the combination $2 \Delta_3(x)-\Delta_2^{\otimes 2}(x)$ occurring at ${\cal O}(a_s^3)$ in eq.(\ref {KL-x-expand}). Using the above results, one finds:

\begin{eqnarray}\label{2d3-d2d2}
2 \Delta_3(x)-\Delta_2^{\otimes 2}(x)&=&-4 C_F \beta_0\frac{\ln^2(1-x)}{1-x}
+(2\Delta_{L,2}+...)\frac{\ln(1-x)}{1-x}+...\\
&+ &(24 C_F^2+0\times C_F\beta_0+0\times C_FC_A)\ln^2(1-x)
+(-2\Delta_{L,2}+...)\ln(1-x)+...\ ,\nonumber\end{eqnarray}
with

\begin{equation}\Delta_{L,2}=-16[\zeta_3-\zeta_2+(1-\zeta_2)^2]( C_A-2 C_F)^2\ ,\label{DL2-bis}\end{equation}
where the $\zeta_3-\zeta_2$ term arises from $\Delta_3(x)$, and the $(1-\zeta_2)^2$ term  from $\Delta_2^{\otimes 2}(x)$. Eq.(\ref{DL2-bis}) agrees with eq.(\ref{DL2}). One observes that:

\noindent i) \underline{Concerning leading logarithms}: the $C_F\beta_0$ and $C_F C_A$ color factors contributions to the $\ln^2(1-x)$ term have cancelled between $\Delta_3(x)$ and  $\Delta_2^{\otimes 2}(x)$.

\noindent ii)  \underline{Concerning next-to-leading logarithms}:
only the $( C_A-2 C_F)^2$ color factor contribution to the $\ln(1-x)/(1-x)$ and $\ln(1-x)$  terms has been written down in eq.(\ref{2d3-d2d2}), since the other color factors do not have any $\zeta_3$, nor $\zeta_2^2$, terms: in particular, the $\zeta_3$ parts of the $C_F(C_A-2C_F)$ factor  have cancelled between the $\Delta_3(x)$ and the $\Delta_2^{\otimes 2}(x)$ contributions to the  $\ln(1-x)/(1-x)$ coefficient. Again, I note that the $( C_A-2 C_F)^2$ color factor contributions to  the  $\ln(1-x)/(1-x)$ and  $\ln(1-x)$ terms are {\em equal up to a sign}, as follows from a previous remark.

\section{Large-$\beta_0$ results}
The large-$n_f$ (``large-$\beta_0$'') longitudinal coefficient function has been computed to all orders in $\alpha_s$ by various methods \cite{Gracey:1995aj,Mankiewicz:1997gz,Dasgupta:1996hh}. Here  I use  the dispersive approach of \cite{Dasgupta:1996hh}  to derive eq.(\ref{K_L-large-b0}).
The derivation  is not quite trivial and  proceeds in two steps: 1) first one proves the existence of a momentum space  threshold resummation formula  for the coefficient function itself, namely as $x\rightarrow 1$:

\begin{equation} \left.{\cal C}_L(x, Q^2)\right\vert_{\rm large \,\,\beta_0}\, \sim{\cal C}_{L,0}((1-x)Q^2)\ .\label{large-b0-coeff-large-x}
\end{equation}
 2) Next, one shows that eq.(\ref{large-b0-coeff-large-x}) implies eq.(\ref{K_L-large-b0}). 
\subsection{Coefficient function}
To derive eq.(\ref{large-b0-coeff-large-x}), I use a method similar to the one  in \cite{Grunberg:2007nc}, based on 
 a dispersive approach \cite{Beneke:1994qe,Ball:1995ni,Dokshitzer:1995qm,Dasgupta:1996hh}. I first observe  the large-$\beta_0$ longitudinal coefficient function ${\cal C}_L$ itself (at the difference of ${\cal C}_2$) obeys the dispersive representation:

\begin{equation}
  \label{finite-r-C-L}
  \left.{\cal C}_L(x, Q^2)\right\vert_{\rm large \,\,\beta_0}\,
  =4 \,C_F\,\int_0^{\infty}\frac{d\mu^2}{\mu^2} a_V^{Mink}(\mu^2)  \dot{\cal F}_L(\mu^2/Q^2,x)\ ,
\end{equation}
where ${\cal F}_L(\mu^2/Q^2,x)$ is the longitudinal  `characteristic function'   (with $\dot{\cal F}_L\equiv -\mu^2\frac{d{\cal F}_L}{d\mu^2}$), which has been computed in \cite{Dasgupta:1996hh}. In eq.(\ref{finite-r-C-L}) $a_V^{Mink}(\mu^2)$ is a `Minkowkian coupling', related formally (i.e. barring the Landau pole) to the one-loop V-scheme coupling 
\begin{equation}a_s^V(Q^2)=\frac{a_s(Q^2)}{1-\frac{5}{3}\beta_0 a_s(Q^2)}\label{a-V}\ \end{equation}
 by the dispersion relation:

\begin{equation}a_s^V(Q^2)=\int_0^{\infty}\frac{d\mu^2}{Q^2}\frac{a_V^{Mink}(\mu^2)}{(1+\mu^2/Q^2)^2}\label{a-Mink}\ .\end{equation}
Throughout this appendix, $a_s(Q^2)$ is  the one-loop $\overline{\rm MS}$ coupling and $\beta_0=-\frac{2}{3}n_f$ is defined  as the abelian (large-$n_f$) part of the one loop beta function coefficient.
Next, one takes the $x\rightarrow 1$ expansion under the integral (\ref{finite-r-C-L}) with a fixed invariant jet mass $W^2=Q^2(1-x)/x\equiv Q^2\,r$. Starting from the expression for ${\cal F}_L(\epsilon,x)$ (with $\epsilon=\mu^2/Q^2$):

\begin{equation}\label{char-L}{\cal F}_L(\epsilon,x)={\cal F}_L^{(r)}(\epsilon,x)\theta(1-x-\epsilon\,x)\
\end{equation}
where ${\cal F}_L^{(r)}(\epsilon,x)$ is the real gluon emission contribution (there is no  virtual contribution), and using the explicit expression for  ${\cal F}_L^{(r)}(\epsilon,x)$  in \cite{Dasgupta:1996hh},  one obtains
the  small $r$ expansion  (at {\em fixed} $\xi\equiv\frac{\epsilon}{r}=\frac{\mu^2}{W^2}$):
\begin{equation}
\label{r-exp-rchar-L}
{\cal F}_L^{(r)}(\epsilon,x)={\cal F}_{L,0}^{(r)}(\xi)+r\,{\cal F}_{L,1}^{(r)}(\xi)+{\cal O}(r^2)\ ,
\end{equation}
with
\begin{align}
\label{Fr-L-exp}
\begin{split}
{\cal F}_{L,0}^{(r)}(\xi)&=(1-\xi)^2\\
{\cal F}_{L,1}^{(r)}(\xi)&=-(1+4\,\xi\ln \xi+2\,\xi-3\,\xi^2)
\ .
\end{split}
\end{align}
Thus ($^.\equiv -\mu^2\frac{d}{d\mu^2}$):
\begin{equation}
\label{r-exp-drchar-L}
\dot{\cal F}_L^{(r)}(\epsilon,x)=\dot{\cal F}_{L,0}^{(r)}(\xi)+r\,\dot{\cal F}_{L,1}^{(r)}(\xi)+{\cal O}(r^2)\ ,
\end{equation}
with
\begin{align}
\label{dFr-L-exp}
\begin{split}
\dot{\cal F}_{L,0}^{(r)}(\xi)&=2\,\xi-2\,\xi^2\\
\dot{\cal F}_{L,1}^{(r)}(\xi)&=4\,\xi\ln \xi+6\,\xi-6\,\xi^2
\ .
\end{split}
\end{align}
Using these results into eq.(\ref{char-L}), and noting that 

\begin{equation}\label{theta}\theta(1-x-\epsilon\,x)=\theta(\xi<1)\ ,\end{equation} 
 one gets:

\begin{equation}
\label{r-exp-char-L}
{\cal F}_L(\epsilon,x)={\cal F}_{L,0}(\xi)+r\,{\cal F}_{L,1}(\xi)+{\cal O}(r^2)
\end{equation}
with
\begin{align}
\label{F_JL}
\begin{split}
{\cal F}_{L,i}(\xi)={\cal F}_{L,i}^{(r)}(\xi)\,\theta(\xi<1)
\ ,
\end{split}
\end{align}
and also

\begin{equation}
\label{r-exp-dchar-L}
\dot{\cal F}_L(\epsilon,x)=\dot{\cal F}_{L,0}(\xi)+r\,\dot{\cal F}_{L,1}(\xi)+{\cal O}(r^2)
\end{equation}
with
\begin{align}
\label{dF_JL}
\begin{split}
\dot{\cal F}_{L,i}(\xi)&=\dot{\cal F}_{L,i}^{(r)}(\xi)\,\theta(\xi<1)
\ ,
\end{split}
\end{align}
where I used the fact that all the  terms ${\cal F}_{L,i}^{(r)}(\xi)$  in (\ref{r-exp-rchar-L})   vanish at $\xi=1$
(which allows to treat the $\theta$ function effectively as a multiplicative constant when taking the derivative). This feature actually follows from the stronger property that the {\em exact}  function ${\cal F}_L^{(r)}(\epsilon,x)$, as well as its first derivative $\dot{\cal F}_L^{(r)}(\epsilon,x)$, {\em both vanish} at $\xi=1$, i.e. for $\epsilon=\frac{1-x}{x}$. Indeed one finds, expanding around $\xi=1$ at fixed $r$:
\begin{equation}\label{xi-exp-rchar-L}{\cal F}_L^{(r)}(\epsilon,x)=(1+r)\,(1-\xi)^2+{\cal O}\left((1-\xi)^3\right)\ .\end {equation}
Reporting eq.(\ref{r-exp-dchar-L}) into (\ref{finite-r-C-L}), one  thus obtains the small $r$ expansion ($r=(1-x)/x$) of the (large-$\beta_0$) longitudinal coefficient function:

\begin{equation}
\label{r-expansion-CL}
\left.{\cal C}_L(x, Q^2)\right\vert_{\rm large \,\,\beta_0}\,={\cal C}_{L,0}\left(W^2\right)\,+r\,{\cal C}_{L,1}\left(W^2\right)\,+{\cal O}\left(r^2\right)\,
\end{equation}
with ($\xi=\mu^2/W^2$):

\begin{align}
\label{disp-rep-CL}
 \begin{split}
 {\cal C}_{L,i}(W^2)
&=4\,C_F\int_0^{\infty}\frac{d\mu^2}{\mu^2}a_V^{Mink}(\mu^2) \,  \dot{\cal F}_{L,i}(\xi)
\ .
\end{split}
\end{align}
Eq.(\ref{r-expansion-CL}) shows that there is a momentum space  threshold resummation formula valid to {\em all orders} in $1-x$ for the coefficient function itself, whose leading order term proves in particular eq.(\ref{large-b0-coeff-large-x}).

\noindent As a by-product, eq.(\ref{disp-rep-CL})
 also  gives the  dispersive representations of the  ${\cal C}_{L,i}$ functions.
I note  that $\dot{\cal F}_{L,i}(0)=0$ (see eq.(\ref{dFr-L-exp})), as it should for eq.(\ref{disp-rep-CL}) to be infrared finite. This result follows from the  observation  that ${\cal F}_L^{(r)}(\epsilon=0,x)$ is {\em finite}:
\begin{equation}
\label{Fr-L-0}
{\cal F}_L^{(r)}(\epsilon=0,x)=x\ ,
\end{equation}
 which guarantees the vanishing of its logarithmic derivative at $\epsilon=0$ (or equivalently $\xi=0$) for {\em any} $x$ (equivalently any $r$):
\begin{equation}
\label{dFr-L-0}
\dot{\cal F}_L^{(r)}(\epsilon=0,x)=0\ .
\end{equation}
I further note that
   $\dot{\cal F}_{L,0}(\xi)$ is {\em analytic} for $\xi\rightarrow 0$, which implies \cite{Beneke:1994qe,Ball:1995ni,Dokshitzer:1995qm} the {\em leading} function ${\cal C}_{L,0}$  {\em does not } have infrared renormalons. On the other hand, the  vanishing for $\xi\rightarrow 0$ of  $\dot{\cal F}_{L,i}(\xi)$   is {\em logarithmically enhanced}, hence non-analytic, for $i\geq 1$, which implies renormalons  in the subleading   ${\cal C}_{L,i}$ ($i\geq 1$) functions.
   
   Finally, the perturbative expansion in the $\overline{\rm MS}$ scheme  of the ${\cal C}_{L,i}$ functions   can easily be obtained from eq.(\ref{a-V}), (\ref{a-Mink}), (\ref{dFr-L-exp}), (\ref{dF_JL}) and (\ref{disp-rep-CL}). Setting:
   
   \begin{equation}{\cal C}_{L,i}(Q^2)=4C_F\sum_{j=1}^\infty (\beta_0)^{j-1} c_{L,i}^{(j)}\  a_s^{j}\label{CLi-exp}\ ,
\end{equation}
 one finds (e.g. for  the first five orders of the leading $i=0$ function):

\begin{eqnarray}\label{large-b0-coeff}c_{L,0}^{(1)}&=&1\nonumber\\
c_{L,0}^{(2)}&=&\frac{19}{6}\nonumber\\
c_{L,0}^{(3)}&=&\frac{203}{18}-2\zeta_2\nonumber\\
c_{L,0}^{(4)}&=&\frac{4955}{108}-19\zeta_2\nonumber\\
c_{L,0}^{(5)}&=&\frac{34883}{162}-\frac{406}{3}\zeta_2+\frac{36}{5}\zeta_2^2\ .
\end{eqnarray}
We shall also need  the moment space version of eq.(\ref{large-b0-coeff-large-x}). Taking the moments, one gets at large $N$:

\begin{equation}\label{momentspace-coeff-largeb0} \left.{\widetilde {\cal C}}_L(N,a_s)\right\vert_{\rm large \,\,\beta_0}\sim\int_0^1dx\, x^{N-1} {\cal C}_{L,0}((1-x)Q^2)\ . \end{equation} 
 Renormalization group invariance at large-$\beta_0$ implies:

\begin{eqnarray}\label{RG-largeb0} {\cal C}_{L,0}((1-x)Q^2)&=&4C_F\Big[a_s+a_s^2\ \beta_0(-L_x+c_{L,0}^{(2)})+a_s^3\ \beta_0^2(L_x^2-2 c_{L,0}^{(2)}L_x+c_{L,0}^{(3)})\nonumber\\
& &+a_s^4\ \beta_0^3(-L_x^3+3 c_{L,0}^{(2)}L_x^2-3 c_{L,0}^{(3)}L_x+c_{L,0}^{(4)})\\
& &+a_s^5\ \beta_0^4(L_x^4-4 c_{L,0}^{(2)}L_x^3+6 c_{L,0}^{(3)}L_x^2-4c_{L,0}^{(4)}L_x+c_{L,0}^{(5)})+{\cal O}(a_s^6)\Big]\nonumber\end{eqnarray}
where $L_x=\ln(1-x)$. Using eq.(\ref{RG-largeb0}) into (\ref{momentspace-coeff-largeb0}) then yields, for $N\rightarrow\infty$:

\begin{eqnarray}\label{momentspace-coeff-largeb0-bis}  \left.{\widetilde {\cal C}}_L(N,a_s)\right\vert_{\rm large \,\,\beta_0}&\sim&\frac{4C_F}{N}\Big[a_s+a_s^2\ \beta_0\Big({\tilde L}+c_{L,0}^{(2)}\Big)+a_s^3\ \beta_0^2\Big({\tilde L}^2+2 c_{L,0}^{(2)}{\tilde L}+c_{L,0}^{(3)}+\zeta_2\Big)\nonumber\\
& &+a_s^4\ \beta_0^3\Big({\tilde L}^3+3 c_{L,0}^{(2)}{\tilde L}^2+(3 c_{L,0}^{(3)}+3\zeta_2){\tilde L}+c_{L,0}^{(4)}+3\zeta_2 c_{L,0}^{(2)}+2\zeta_3\Big)\nonumber\\
& &+a_s^5\ \beta_0^4\Big({\tilde L}^4+4 c_{L,0}^{(2)}{\tilde L}^3+(6 c_{L,0}^{(3)}+6\zeta_2){\tilde L}^2+(4c_{L,0}^{(4)}+12\zeta_2 c_{L,0}^{(2)}+8\zeta_3){\tilde L}\nonumber\\
& &+c_{L,0}^{(5)}+6\zeta_2 c_{L,0}^{(3)}+8\zeta_3 c_{L,0}^{(2)}+\frac{27}{5}\zeta_2^2\Big)+{\cal O}(a_s^6)\Big]\end{eqnarray}
where ${\tilde L}=\ln N+\gamma_E$. Eq.(\ref{large-b0-coeff}) and (\ref{momentspace-coeff-largeb0-bis}) can be checked more directly using the results in \cite{Mankiewicz:1997gz}. However, the advantage of the dispersive approach used here is that it gives a proof of eq.(\ref{large-b0-coeff-large-x}) to all orders in $a_s$.

\subsection{Physical evolution kernel}
To show eq.(\ref{large-b0-coeff-large-x}) implies eq.(\ref{K_L-large-b0}) one proceeds again in two steps: 1) first, one proves that eq.(\ref{large-b0-coeff-large-x}) implies in moment space for $N\rightarrow\infty$:

\begin{equation}\label{moment-K_L-large-b0}
\left.{\widetilde K}_L(Q^2,N)\right\vert_{\rm large \,\,\beta_0}\,\sim N\int_0^1 dx\ x^{N-1}\ {\cal B}_{\infty}((1-x)Q^2)\end{equation}
where ${\cal B}_{\infty}(Q^2)$ is a large-$\beta_0$ renormalization group invariant quantity similar to ${\cal C}_{L,0}(Q^2)$,  which satisfies an analogue of eq.(\ref{RG-largeb0}):

\begin{eqnarray}\label{RG-largeb0-B} {\cal B}_{\infty}((1-x)Q^2)&=&-\beta_0\Big[a_s+a_s^2\ \beta_0(-L_x+b_{\infty}^{(2)})+a_s^3\ \beta_0^2(L_x^2-2 b_{\infty}^{(2)}L_x+b_{\infty}^{(3)})\nonumber\\
& &+a_s^4\ \beta_0^3(-L_x^3+3 b_{\infty}^{(2)}L_x^2-3 b_{\infty}^{(3)}L_x+b_{\infty}^{(4)})\\
& &+a_s^5\ \beta_0^4(L_x^4-4 b_{\infty}^{(2)}L_x^3+6 b_{\infty}^{(3)}L_x^2-4b_{\infty}^{(4)}L_x+b_{\infty}^{(5)})+{\cal O}(a_s^6)\Big]\nonumber\ .\end{eqnarray}
2) Next, one shows that eq.(\ref{moment-K_L-large-b0}) implies eq.(\ref{K_L-large-b0}).

1) One notes that eq.(\ref{moment-K_L-large-b0}) and (\ref{RG-largeb0-B} ) imply for $N\rightarrow\infty$ (similarly to eq.(\ref{momentspace-coeff-largeb0-bis})):

\begin{eqnarray}\label{momentspace-KL-largeb0-bis}  \left.{\widetilde K}_L(Q^2,N)\right\vert_{\rm large \,\,\beta_0}&\sim&-\beta_0\Big[a_s+a_s^2\ \beta_0\Big({\tilde L}+b_{\infty}^{(2)}\Big)+a_s^3\ \beta_0^2\Big({\tilde L}^2+2 b_{\infty}^{(2)}{\tilde L}+b_{\infty}^{(3)}+\zeta_2\Big)\nonumber\\
& &+a_s^4\ \beta_0^3\Big({\tilde L}^3+3 b_{\infty}^{(2)}{\tilde L}^2+(3 b_{\infty}^{(3)}+3\zeta_2){\tilde L}+b_{\infty}^{(4)}+3\zeta_2 b_{\infty}^{(2)}+2\zeta_3\Big)\nonumber\\
& &+a_s^5\ \beta_0^4\Big({\tilde L}^4+4 b_{\infty}^{(2)}{\tilde L}^3+(6 b_{\infty}^{(3)}+6\zeta_2){\tilde L}^2+(4b_{\infty}^{(4)}+12\zeta_2 b_{\infty}^{(2)}+8\zeta_3){\tilde L}\nonumber\\
& &+b_{\infty}^{(5)}+6\zeta_2 b_{\infty}^{(3)}+8\zeta_3 b_{\infty}^{(2)}+\frac{27}{5}\zeta_2^2\Big)+{\cal O}(a_s^6)\Big]\ .\end{eqnarray}
On the other hand, eq.(\ref{K-tilde2}) yields at large-$\beta_0$ (where the anomalous dimension term can be neglected):

\begin{equation} \left.{\widetilde K}_L(Q^2,N)\right\vert_{\rm large \,\,\beta_0}=-\beta_0\ a_s^2\frac{d\ln  \left.{\widetilde {\cal C}}_L(N,a_s)\right\vert_{\rm large \,\,\beta_0}}{da_s}\label{K-tilde2-largeb0}\ .\end{equation}
Using eq.(\ref{momentspace-coeff-largeb0-bis}) into eq.(\ref{K-tilde2-largeb0}), one then finds that the latter is indeed consistent with eq.(\ref{momentspace-KL-largeb0-bis}), and determines:

\begin{eqnarray}b_{\infty}^{(2)}&=&c_{L,0}^{(2)}\label{b-large-b0}\\
b_{\infty}^{(3)}&=&2 c_{L,0}^{(3)}-(c_{L,0}^{(2)})^2+\zeta_2\nonumber\\
b_{\infty}^{(4)}&=&3 c_{L,0}^{(4)}+(c_{L,0}^{(2)})^3-3c_{L,0}^{(2)} c_{L,0}^{(3)}+3\zeta_2 c_{L,0}^{(2)}+4\zeta_3\nonumber\\
b_{\infty}^{(5)}&=&4 c_{L,0}^{(5)}-(c_{L,0}^{(2)})^4+4(c_{L,0}^{(2)})^2 c_{L,0}^{(3)}-2(c_{L,0}^{(3)})^2-4c_{L,0}^{(2)} c_{L,0}^{(4)}+8\zeta_2 c_{L,0}^{(3)}-2\zeta_2 (c_{L,0}^{(2)})^2+16\zeta_3 c_{L,0}^{(2)}+\frac{41}{5}\zeta_2^2\nonumber\ ,\end{eqnarray}
which checks eq.(\ref{moment-K_L-large-b0}) up to ${\cal O}(a_s^5)$ order (the check can be clearly performed to arbitrary high order).

2) To show that eq.(\ref{moment-K_L-large-b0}) implies eq.(\ref{K_L-large-b0}), it is convenient to introduce the formal (i.e. infrared divergent at $x=1$) Mellin moment  integral:

\begin{equation}\label{moment-integral}
E(Q^2,N)\equiv \int_0^1 dx\ x^{N-1}\frac{1}{r}\ {\cal B}_{\infty}(W^2)\end{equation}
where $W^2=rQ^2$ with $r=(1-x)/x$. One first observes that $dE/d\ln Q^2$ is {\em finite}. Indeed, eq.(\ref{moment-integral}) can be written equivalently as:

\begin{eqnarray}\label{moment-integral-bis}
E(Q^2,N)&=& \int_0^1 dx\ \Big[x^{N-1}\frac{1}{r}\ {\cal B}_{\infty}(W^2)-\frac{1}{1-x}\ {\cal B}_{\infty}((1-x)Q^2)\Big]+\int_0^{Q^2}\frac{dk^2}{k^2}{\cal B}_{\infty}(k^2)\nonumber\\
&\equiv&\int_0^1 dx\ x^{N-1}\Big[\frac{1}{r}\ {\cal B}_{\infty}(W^2)\Big]_++\int_0^{Q^2}\frac{dk^2}{k^2}{\cal B}_{\infty}(k^2)\end{eqnarray}
where the first integral on the right hand side is finite, and the infrared divergence is entirely contained in the second integral. Taking the derivative one thus gets:

\begin{equation}\label{dmoment-integral-bis}
\frac{dE(Q^2,N)}{d\ln Q^2}=\int_0^1 dx\ x^{N-1}\Big[\frac{1}{r}\ \frac{d{\cal B}_{\infty}(W^2)}{d\ln Q^2}\Big]_++{\cal B}_{\infty}(Q^2)\end{equation}
where the right hand side is finite. Eq.(\ref{dmoment-integral-bis}) also shows that $dE(Q^2,N)/d\ln Q^2$ is the Mellin moment of $\Big[\frac{1}{r}\ \frac{d{\cal B}_{\infty}(W^2)}{d\ln Q^2}\Big]_++{\cal B}_{\infty}(Q^2)\delta(1-x)$.

On the other hand, performing the change of variable $r=k^2/Q^2$  in the integral of eq.(\ref{moment-integral}), one gets:

\begin{equation}\label{moment-integral-ter}
E(Q^2,N)\equiv \int_0^{\infty} \frac{dk^2}{k^2}\Big(\frac{1}{1+\frac{k^2}{Q^2}}\Big)^{N+1}{\cal B}_{\infty}(k^2)\ ,\end{equation}
which yields:

\begin{equation}\label{dmoment-integral-ter}
\frac{dE(Q^2,N)}{d\ln Q^2}=\frac{N+1}{Q^2}\int_0^{\infty} dk^2\Big(\frac{1}{1+\frac{k^2}{Q^2}}\Big)^{N+2}{\cal B}_{\infty}(k^2)=\frac{N+1}{N}\int_0^{\infty} dt\Big(\frac{1}{1+\frac{t}{N}}\Big)^{N+2}{\cal B}_{\infty}(tQ^2/N)\ ,\end{equation}
where the change of variable $t=Nk^2/Q^2$ has been performed. Moreover at large $N$, eq.(\ref{moment-K_L-large-b0}) is equivalent to:

\begin{equation}\label{moment-K_L-large-b0-bis}
\left.{\widetilde K}_L(Q^2,N)\right\vert_{\rm large \,\,\beta_0}\,\sim
N\int_0^1 dx\ x^{N-1}\ {\cal B}_{L,\infty}^{\delta}(W^2)\equiv\int_0^{\infty} dt\Big(\frac{1}{1+\frac{t}{N}}\Big)^{N+1}{\cal B}_{\infty}(tQ^2/N)\ ,\end{equation}
where  $t=Nr$. For $N\rightarrow\infty$ with $Q^2/N$ fixed, the right hand sides of eq.(\ref{dmoment-integral-ter}) and (\ref{moment-K_L-large-b0-bis}) are both equivalent to $\int_0^{\infty} dt\exp(-t)\  {\cal B}_{\infty}(tQ^2/N)$. Thus at large $N$ we have, using eq.(\ref{dmoment-integral-bis}):

\begin{eqnarray}\label{moment-K_L-large-b0-ter}
\left.{\widetilde K}_L(Q^2,N)\right\vert_{\rm large \,\,\beta_0}&\sim&N\int_0^1 dx\ x^{N-1}\ {\cal B}_{\infty}((1-x)Q^2)\\
&\sim&\frac{dE(Q^2,N)}{d\ln Q^2}\sim\int_0^1 dx\ \frac {x^{N-1}-1}{1-x}\ \frac{d{\cal B}_{\infty}((1-x)Q^2)}{d\ln Q^2}+{\cal B}_{\infty}(Q^2)\nonumber\ ,\end{eqnarray}
which proves eq.(\ref{K_L-large-b0}) by inverting back to momentum space. The  large--$N$ equivalence of the two expressions on the right hand side of eq.(\ref{moment-K_L-large-b0-ter}) can easily be checked order by order in $a_s$ using eq.(\ref{RG-largeb0-B}).

\noindent Finally, I note it is likely that eq.(\ref{K_L-large-b0}) can be generalized to all orders in $(1-x)$, as a consequence of eq.(\ref{r-expansion-CL}):

\begin{equation}
\label{r-expansion-KL}
\left. K_L(x, Q^2)\right\vert_{\rm large \,\,\beta_0}\,=\Big[\frac{1}{r}\ \frac{d{\cal B}_{\infty}(W^2)}{d\ln Q^2}\Big]_++{\cal B}_{\infty}(Q^2)\delta(1-x)+\left.
{\cal J}_{0L}\left(W^2\right)\right\vert_{\rm large \,\,\beta_0}\,+r\left.
{\cal J}_{1L}\left(W^2\right)\right\vert_{\rm large \,\,\beta_0}\,+{\cal O}\left(r^2\right)\ .
\end{equation}


\begin{thebibliography}{99}


\bibitem{Sterman:1986aj}
  G.~Sterman,
  {\em Nucl. Phys.}  {\bf B281} (1987) 310.

\bibitem{Catani:1989ne}
  S.~Catani and L.~Trentadue,
  {\em Nucl. Phys.}  {\bf B327} (1989) 323.


\bibitem{Akhoury:1998gs}
  R.~Akhoury, M.~G.~Sotiropoulos and G.~Sterman,
  Phys.\ Rev.\ Lett.\  {\bf 81} (1998) 3819
  [arXiv:hep-ph/9807330].

\bibitem{Sotiropoulos:1999hy}
  M.~G.~Sotiropoulos, R.~Akhoury and G.~Sterman,
  arXiv:hep-ph/9903442.

\bibitem{Akhoury:2003fw}
  R.~Akhoury and M.~G.~Sotiropoulos,
  arXiv:hep-ph/0304131.





\bibitem{Kramer:1996iq}
  M.~Kramer, E.~Laenen and M.~Spira,
  Nucl.\ Phys.\  B {\bf 511} (1998) 523
  [arXiv:hep-ph/9611272].

\bibitem{Grunberg:2007nc}
  G.~Grunberg,
  arXiv:0710.5693 [hep-ph].


\bibitem{Laenen:2008ux}
  E.~Laenen, L.~Magnea and G.~Stavenga,
  Phys.\ Lett.\  B {\bf 669} (2008) 173
  [arXiv:0807.4412 [hep-ph]].

\bibitem{Laenen:2008gt}
  E.~Laenen, G.~Stavenga and C.~D.~White,
  arXiv:0811.2067 [hep-ph].

\bibitem{Moch:2009mu}
  S.~Moch and A.~Vogt,
  JHEP {\bf 0904} (2009) 081
  [arXiv:0902.2342 [hep-ph]].

\bibitem{Grunberg:2009yi}
  G.~Grunberg and V.~Ravindran,
  arXiv:0902.2702 [hep-ph] (to be published in JHEP).

\bibitem{Moch:2009hr}
  S.~Moch and A.~Vogt,
  arXiv:0909.2124 [hep-ph].

\bibitem{Furmanski:1981cw}
  W.~Furmanski and R.~Petronzio,
  Z.\ Phys.\  C {\bf 11} (1982) 293.



\bibitem{Grunberg:1982fw}
  G.~Grunberg,
  {\em Phys. Rev.} {\bf D29} (1984) 2315.


\bibitem{Catani:1996sc}
  S.~Catani,
  Z. Phys. {\bf C75} (1997) 665
  [hep-ph/9609263].

\bibitem{Blumlein:2000wh}
  J.~Blumlein, V.~Ravindran and W.~L.~van Neerven,
  Nucl.\ Phys.\  B {\bf 586} (2000) 349
  [arXiv:hep-ph/0004172].





\bibitem{vanNeerven:2001pe}
  W.~L.~van Neerven and A.~Vogt,
  Nucl.\ Phys.\  B {\bf 603} (2001) 42
  [arXiv:hep-ph/0103123].

\bibitem{Moch:2004xu}
  S.~Moch, J.~A.~M.~Vermaseren and A.~Vogt,
  Phys.\ Lett.\  B {\bf 606} (2005) 123
  [arXiv:hep-ph/0411112].

\bibitem{Vermaseren:2005qc}
J.~A.~M.~Vermaseren,  A.~Vogt and S.~Moch,
  {\em Nucl. Phys.} {\bf B724} (2005) 3
  [hep-ph/0504242].










\bibitem{Gardi:2007ma}
  E.~Gardi and G.~Grunberg,
  Nucl.\ Phys.\  B {\bf 794} (2008) 61
  [arXiv:0709.2877 [hep-ph]].

\bibitem{Korchemsky:1993uz}
  G.~P.~Korchemsky and G.~Marchesini,
  Phys.\ Lett.\  B {\bf 313} (1993) 433.



\bibitem{Korchemsky:1988si}
  G.~P.~Korchemsky,
  Mod.\ Phys.\ Lett.\  A {\bf 4} (1989) 1257.

\bibitem{Kodaira:1981nh}
  J.~Kodaira and L.~Trentadue,
  Phys.\ Lett.\  B {\bf 112} (1982) 66.



\bibitem{Moch:2004pa}
  S.~Moch, J.~A.~M.~Vermaseren and A.~Vogt,
  {\em Nucl. Phys.} {\bf B688} (2004) 101
  [hep-ph/0403192].

\bibitem{SanchezGuillen:1990iq}
  J.~Sanchez Guillen, J.~Miramontes, M.~Miramontes, G.~Parente and O.~A.~Sampayo,
  Nucl.\ Phys.\  B {\bf 353} (1991) 337.



\bibitem{vanNeerven:1991nn}
  W.~L.~van Neerven and E.~B.~Zijlstra,
  Phys.\ Lett.\  B {\bf 272} (1991) 127.

\bibitem{Moch:1999eb}
  S.~Moch and J.~A.~M.~Vermaseren,
  Nucl.\ Phys.\  B {\bf 573} (2000) 853
  [arXiv:hep-ph/9912355].

\bibitem{Floratos:1977au}
  E.~G.~Floratos, D.~A.~Ross and C.~T.~Sachrajda,
  Nucl.\ Phys.\  B {\bf 129} (1977) 66
  [Erratum-ibid.\  B {\bf 139} (1978) 545].



\bibitem{Curci:1980uw}
  G.~Curci, W.~Furmanski and R.~Petronzio,
  Nucl.\ Phys.\  B {\bf 175} (1980) 27.





\bibitem{Dokshitzer:2005bf}
  Yu.~L.~Dokshitzer, G.~Marchesini and G.~P.~Salam,
  Phys.\ Lett.\  B {\bf 634} (2006) 504
  [arXiv:hep-ph/0511302].
  
\bibitem{Basso:2006nk}
  B.~Basso and G.~P.~Korchemsky,
  Nucl.\ Phys.\  B {\bf 775} (2007) 1
  [arXiv:hep-th/0612247].


\bibitem{Vogt}
A.~Vogt, private communication.


\bibitem{Gracey:1995aj}
  J.~A.~Gracey,
  arXiv:hep-ph/9509276.



\bibitem{Dasgupta:1996hh}
  M.~Dasgupta and B.~R.~Webber,
  Phys.\ Lett.\  B {\bf 382} (1996) 273
  [arXiv:hep-ph/9604388].

\bibitem{Mankiewicz:1997gz}
  L.~Mankiewicz, M.~Maul and E.~Stein,
  Phys.\ Lett.\  B {\bf 404} (1997) 345
  [arXiv:hep-ph/9703356].

\bibitem{Beneke:1994qe}
  M.~Beneke and V.~M.~Braun,
  Phys.\ Lett.\  B {\bf 348} (1995) 513
  [arXiv:hep-ph/9411229].

\bibitem{Ball:1995ni}
  P.~Ball, M.~Beneke and V.~M.~Braun,
  Nucl.\ Phys.\  B {\bf 452} (1995) 563
  [arXiv:hep-ph/9502300].

\bibitem{Dokshitzer:1995qm}
  Y.~L.~Dokshitzer, G.~Marchesini and B.~R.~Webber,
  Nucl.\ Phys.\  B {\bf 469} (1996) 93
  [arXiv:hep-ph/9512336].




\end{thebibliography}
\end{document}